\documentclass[review]{elsarticle}

\usepackage{lineno,hyperref}
\modulolinenumbers[5]
\usepackage{hyperref}

\journal{Materials Science in Semiconductor Processing}

\usepackage{url}  
\usepackage{graphicx} 
\usepackage{revsymb}
\usepackage{bm}
\usepackage{textcomp}
\usepackage{revsymb}
\usepackage{amssymb}
\usepackage{ifsym}
\usepackage{wasysym}
\usepackage{txfonts}
\usepackage{tipa}

\begin{document}

\begin{frontmatter}
	 



\title{Diffusion of hydrogen atoms in silicon layers deposited from molecular beams on dielectric substrates\tnoteref{Grants}}

\author[GPI]{Kirill~V.~Chizh\corref{Yuryev}} 
\ead{chizh@kapella.gpi.ru}

\author[GPI]{Larisa~V.~Arapkina}
\ead{arapkina@kapella.gpi.ru}

\author[GPI]{Dmitry~B.~Stavrovsky}
\ead{stavr@nsc.gpi.ru} 

\author[BGU]{Peter~I.~Gaiduk}
\ead{gaiduk@bsu.by}

\author[GPI]{Vladimir~A.~Yuryev\corref{Yuryev}}
\cortext[Yuryev]{Corresponding author}
\ead{vyuryev@kapella.gpi.ru}

\address[GPI]{Prokhorov General Physics Institute of the Russian Academy of Sciences,\\38 Vavilov Street, 119991 Moscow, Russia}
\address[BGU]{Department of Physical Electronics and Nanotechnology, Belarusian State University,\\4 Prospekt Nezavisimosti, 220030 Minsk, Belarus}

\tnotetext[Grants]{The research was funded by the Russian Foundation for Basic Research 
	[grant number 18-52-00033]
	and 
	the Belarusian Republican Foundation for Fundamental Research [grant number T18P-190].}

\date{}

\begin{abstract}
In the paper, the processes occurring during low-temperature growth of non-hydrogenated amorphous Si and polycrystalline Si films on multilayer Si$_3$N$_4$/SiO$_2$/c-Si substrates from molecular beams under conditions of ultrahigh vacuum are studied in detail.
Diffusion of hydrogen atoms from a dielectric layer into the growing film is shown to accompany the growth of a silicon film on a Si$_3$N$_4$ layer deposited by CVD or on a SiO$_2$ layer obtained by thermal oxidation of a silicon wafer.
The process of hydrogen migration from the dielectric substrates into the silicon film is studied using FTIR spectroscopy.
The reduction of IR absorption at the bands related to the N--H bonds vibrations and the increase of IR absorption at the bands relating to the Si--N bonds vibrations in IR spectra demonstrate that hydrogen atoms leave Si$_3$N$_4$ layer during Si deposition from a molecular beam. 
The absorption band assigned to the valence vibrations of the Si--H bond at $\sim 2100$~cm$^{-1}$ emerging in IR spectra obtained at samples deposited both on Si$_3$N$_4$ and SiO$_2$ layers indicates the accumulation of hydrogen atoms in silicon films.
The difference in chemical potentials of hydrogen atoms in the dielectric layer and the silicon film explains the transfer of hydrogen atoms from the Si$_3$N$_4$ or SiO$_2$ layer into the growing silicon film.
\end{abstract}

\begin{keyword}
\texttt{Hydrogen diffusion\sep polycrystalline silicon \sep amorphous silicon \sep molecular beam deposition \sep silicon on dielectric}
\end{keyword}

\end{frontmatter}



\newpage

\section{\label{intro}Introduction}

Presently, MEMS technology is a rapidly developing branch of science, one of the promising areas of which is the production of IR sensors based on amorphous ($\alpha$-Si) or polycrystalline (poly-Si) silicon films formed on Si$_3$N$_4$/SiO$_2$ dielectric layers \cite{PtSi-bolometer,SPIE_PtSi-bolometer-cite,NiSi-bolometer}.
The properties of $\alpha$-Si and poly-Si films are determined both by the technological modes of deposition and by the properties of the substrate, primarily, the upper layer of silicon nitride of a complex Si$_3$N$_4$/SiO$_2$/c-Si substrate.
The most commonly used methods of depositing Si$_3$N$_4$ are LPCVD with thermal and PECVD with plasma-chemical decomposition of nitrogen-containing components.
Application of any of these methods leads to the production of films saturated with hydrogen atoms.
The concentration of hydrogen atoms in PECVD films can reach 25\%, and in LPCVD films 8\% \cite{H_Si3N4,H_Si3N4_nuclear-recoil,H_Si3N4_bonding,H_Si-oxynitride}.
Such hydrogen concentrations can significantly affect the properties of $\alpha$-Si and poly-Si films grown on the surface of a complex Si$_3$N$_4$/SiO$_2$/c-Si substrate due to desorption of hydrogen from the Si$_3$N$_4$ film during and after the deposition of silicon layers.
The protective layer of Si$_3$N$_4$ is known to facilitate a decrease in the recombination of non-equilibrium charge carriers and can reduce the concentration of the majority carriers in the near-surface region of the Si layer due to the formation of a built-in positive charge \cite{Dielectric_surface_passivation}.
Hydrogen atoms are able to diffuse from the Si$_3$N$_4$ layer and passivate the broken bonds at the interface with the Si layer, thereby improving the electrophysical properties of the material \cite{SiN_surface_passivation,Surface_recombination_velocity_SiN-SiO2,c-Si_solar_cells_passivation,Passivation_ultrathin_SiO2,c-Si_surface_passivation}.
At present, protective coatings of Si$_3$N$_4$ deposited by low-temperature PECVD are widely used as a source of hydrogen atoms to prevent degradation of electrophysical properties of solar cells based on single crystalline (c-Si), multicrystalline and amorphous silicon \cite{Passivation_SiN_PECVD}.
However, the study of hydrogen diffusion in such structures is complicated by the fact that the Si$_3$N$_4$ and $\alpha$-Si (poly-Si) layers are hydrogenated and the hydrogen atoms can freely move between the layers.

In this paper, the processes occurring during low-temperature growth from molecular beams under conditions of ultrahigh vacuum (UHV) of non-hydrogenated $\alpha$-Si and poly-Si films on a multilayer Si$_3$N$_4$/SiO$_2$/c-Si substrate are studied in detail.
The Si$_3$N$_4$ layer is the source of hydrogen atoms in our experiment.

Previously, the diffusion of hydrogen atoms into $\alpha$-Si and poly-Si layers from an external source was studied, e.g., in the works \cite{Hydrogenation_kinetics_A-Si,Passivation_grain_boundary,H_passivation_poly-Si,H_loss,Infrared_study_H}.
At low process temperatures ranging from 250 to 450{\textcelsius}, diffusion of hydrogen atoms with an activation energy from $\sim 0.2$ to $\sim 0.4$~eV was observed in those works.
Hydrogen atoms first passivated the broken silicon bonds, and when the temperature was raised above 300{\textcelsius}, a rupture in the weak Si--Si bonds was observed.
The study of the saturation of silicon layers with hydrogen atoms during deposition of Si$_3$N$_4$ layers using PECVD showed that hydrogen atoms penetrated into c-Si to a depth from 10 to 20~nm at temperatures from 500 to 900{\textcelsius}, and pre-treatment of the c-Si surface in NH$_3$ plasma increased the concentration of hydrogen atoms passing into c-Si \cite{H_loss,Infrared_study_H}.

\section{\label{experiment}Sample preparation, experimental methods and equipment  } 

\subsection{Samples \label{samples}}

Experimental samples were prepared by depositing Si layers on different substrates. 
There were three types of substrates used in the experiments: Type~N, Type~O and Type~S. 

Type~N substrates, Si$_3$N$_4$/SiO$_2$/Si(100), were produced as follows.
First, finished on both sides 450~{\textmu}m thick (100)-oriented Czochralski grown (CZ) boron doped silicon wafers ($p$-type, $\varrho$ = 12 $\Omega\,$cm) were cleaned using the RCA process \cite{RCA,Kern_cleaning_evolution,cleaning_handbook}.
Then, 530 nm thick SiO$_2$ layers were formed on both sides of silicon wafers using thermal oxidation. 
This process was performed in three consecutive steps: at first, the wafers were treated in dry oxygen for 30 min, then in wet oxygen for 60 min, and finally, in dry oxygen for 30 min.
A 175 nm thick Si$_3$N$_4$ layer was deposited on both sides of the oxidized wafers by pyrolysis of a monosilane-ammonia mixture at the temperature of 750\,{\textcelsius} for 60~min \cite{PtSi-bolometer,SPIE_PtSi-bolometer-cite,SPIE_SiGe_poly-Si_PtSi}. 

On Type~O substrates, SiO$_2$/Si(111), only 700 nm thick SiO$_2$ film was formed by thermal oxidation of CZ~Si(111):P ($n$-type, $\varrho$ = 100 $\Omega\,$cm) wafers.
Type~S substrates were $p$-type CZ~Si(100):B wafers, $\varrho$ = 12 $\Omega\,$cm.

Silicon films were deposited on the substrates from molecular beams (MB) in the ultra-high vacuum EVA~32 (Riber) molecular-beam epitaxy (MBE) chamber using a solid source with Si evaporation by an electron beam from a high resistivity float zone ingot. 

The MBE chamber was evacuated down to about $10^{-11}$~Torr before the processes.
The pressure did not exceed $5\times10^{-9}$~Torr during Si deposition. 
The deposition rate and coverage were measured using the Inficon  XTC751-001-G1 (Leybold-Heraeus) film thickness monitor  equipped with the graduated
in advance quartz sensors installed in the MBE chamber.
Si deposition rate was $\sim 0.3$~\AA/s; the film growth temperature was varied form 30 to 650{\textcelsius} in different processes for samples produced on Type~N and Type~O substrates.
The thickness of the Si layer was 200~nm.

Before depositing of silicon nitride or silicon, the substrates of all types were cleaned using an identical process that we routinely apply for obtaining clean Si surfaces \cite{phase_transition,Nucleation_high-temperatures,VCIAN2011}. 
At first, they were washed in the ammonia-peroxide solution 
\linebreak 
(NH$_4$OH\,(27\%)\,:\,H$_2$O$_2$\,(30\%)\,:\,H$_2$O [1\,:\,1\,:\,3], 
boiling for 10 min), then rinsed in deionized water, boiled in high-purity isopropyl alcohol ([C$_3$H$_7$OH] $>99.8\,$wt\%, $T\approx 70$\textcelsius), and dried in the isopropyl alcohol vapor (for 10 min) and the clean air.
Additionally, before moving into the MBE chamber, the substrates were annealed at {600\textcelsius} at the residual gas pressure of less than $5 \times 10^{-9}$~Torr in the preliminary
annealing chamber for 6 hours \cite{PtSi-bolometer,SPIE_PtSi-bolometer-cite,SPIE_SiGe_poly-Si_PtSi,stm-rheed-EMRS}. 

Type~S substrates were additionally deoxidized in the MBE chamber at the temperature of {800\textcelsius} at a flux of Si atoms. 
The Si deposition rate during the deoxidation process was  $< 0.1\,${\AA}/s; 
the measured Si coverage was about {30~\AA} \cite{phase_transition}.
The pressure in the MBE chamber increased to nearly $2\times 10^{-9}$~Torr at most during this process.
RH20 reflection high-energy electron diffraction (RHEED) tool (Staib Instruments) installed in the MBE chamber was used for deoxidation monitoring.

The growth of Si films on Type~S substrates was conducted in two stages. First, a 100 nm thick Si layer was deposited at {650\textcelsius} on the wafer surface purified of silicon oxide; then, after sample cooling to the room temperature at the rate of $\sim\,$0.4\,{\textcelsius}/s, a 200~nm thick Si layer was deposited at this temperature.

Samples were heated from the rear side using tantalum radiators in both preliminary annealing and MBE chambers. 
The temperature was monitored with chromel-alumel and tungsten-rhenium thermocouples of the heaters in the preliminary annealing and MBE chambers, respectively. 
The thermocouples were mounted in vacuum near the rear side of the samples. 
They were \textit{in situ} graduated beforehand against the IMPAC IS12-Si infrared pyrometer (LumaSense Technologies)  that measured the sample temperature through MBE chamber windows (for $T >300$\textcelsius;  the graduation curves were  extrapolated to the room temperature for $T <300$\textcelsius). 

The composition of residual atmosphere in the MBE camber was monitored using the RGA-200 residual gas analyzer (Stanford Research Systems) before and during the deposition process.


\subsection{\label{ftir}FTIR analysis}

Fourier transform IR (FTIR) transmission and reflection spectra of the samples were explored using a vacuum IFS-66v/S spectrometer (Bruker). 
The spectral resolution was  10~cm$^{-1}$ that enabled both an adequate recording of all spectral features and filtering away of the high-frequency components related to light interference in silicon wafers. 
An opaque golden mirror was used as reference sample during recording of reflectance spectra.
The instrument was evacuated to the residual air pressure of 2 mbar during spectra recording that enabled a considerable reduction of spectral interferences associated with carbon dioxide and water vapor. 
Direct analysis of the obtained spectra was hampered by interference of probe radiation in thin layers of materials on silicon wafers.
In analyzing the spectra, the transmission spectrum and the reflection spectrum of each sample were summed. 
Such procedure of spectra processing has enabled the reduction of amplitude of baseline variations arisen due to the interference. 
As a result, it has become possible to observe and analyze relatively weak absorption bands.

Additionally, the peak wavenumbers and relative intensities of spectral components were analyzed using the deconvolution procedure. 
We used Gaussian functions for fitting.
Spectra to be compared were normalized to the intensity of the strongest band  assigned to the vibrations of the Si--O bond.

\section{Results and their interpretation \label{results}} 

Both {\it in-situ} structural investigations of the surface of the growing Si layer using RHEED and {\it ex-situ} examinations of the deposited Si layer using high resolution transmission electron microscopy have shown that an amorphous silicon layer forms on Type~N substrates at the temperature range from 30 to {420\textcelsius} \cite{SPIE_SiGe_poly-Si_PtSi,RCEM-2018_PtSi}.
Transition from the amorphous structure of the growing film to the polycrystalline one after the film thickness reaches a critical value---i.e. the formation of a layered poly-Si/amorphous Si film---is observed at the temperature interval from 420 to {500\textcelsius}. 
Finally, only polycrystalline Si layers form at the higher temperatures (for details, see Ref.~\cite{SPIE_SiGe_poly-Si_PtSi}).

\begin{figure}[t]
	\includegraphics*[width=0.5\linewidth]{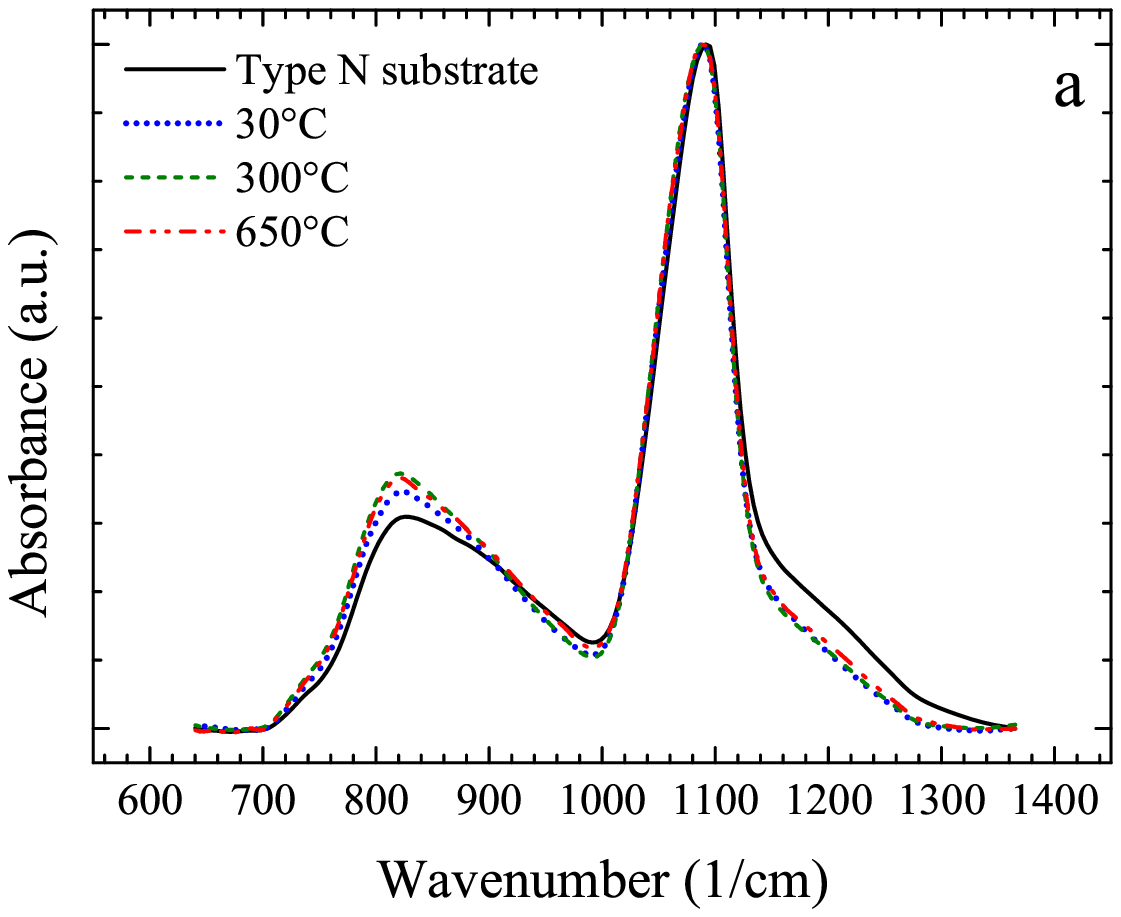}
	\includegraphics*[width=0.5\linewidth]{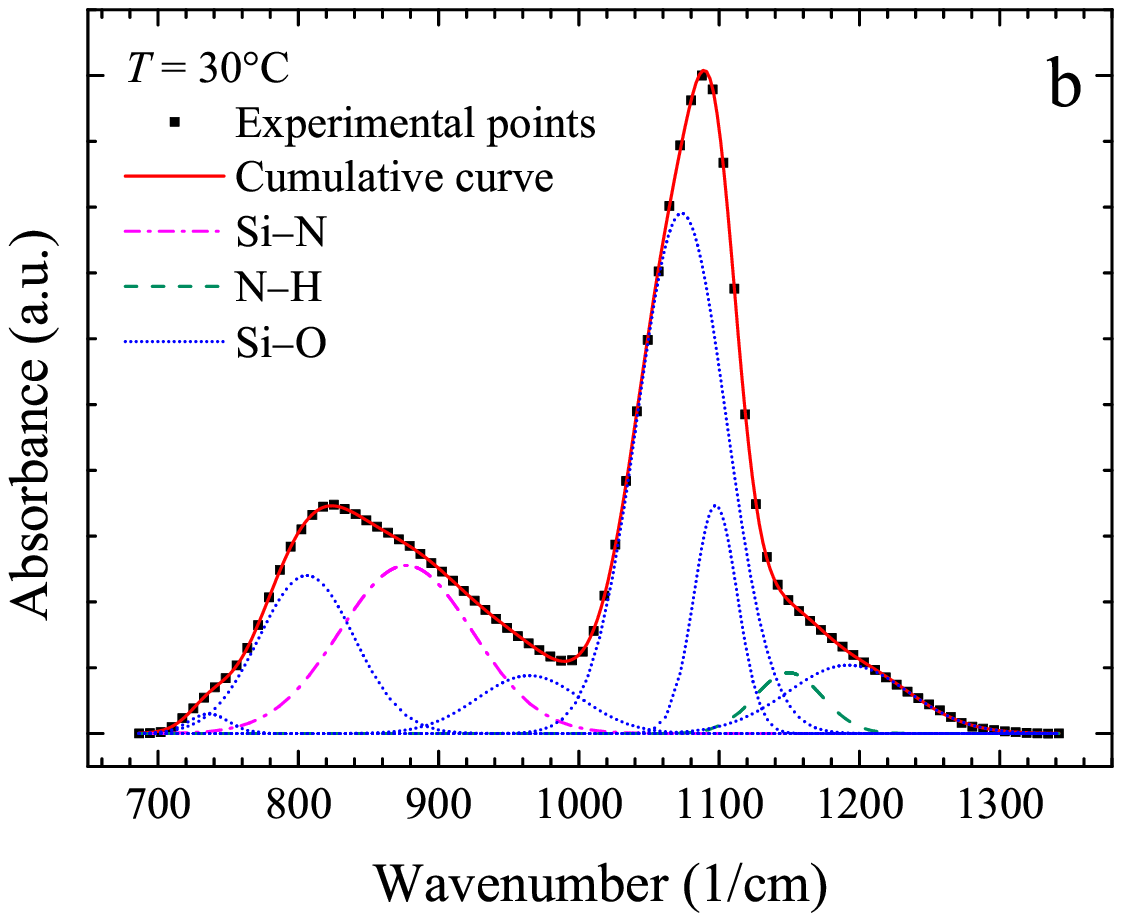}\\
	\includegraphics*[width=0.5\linewidth]{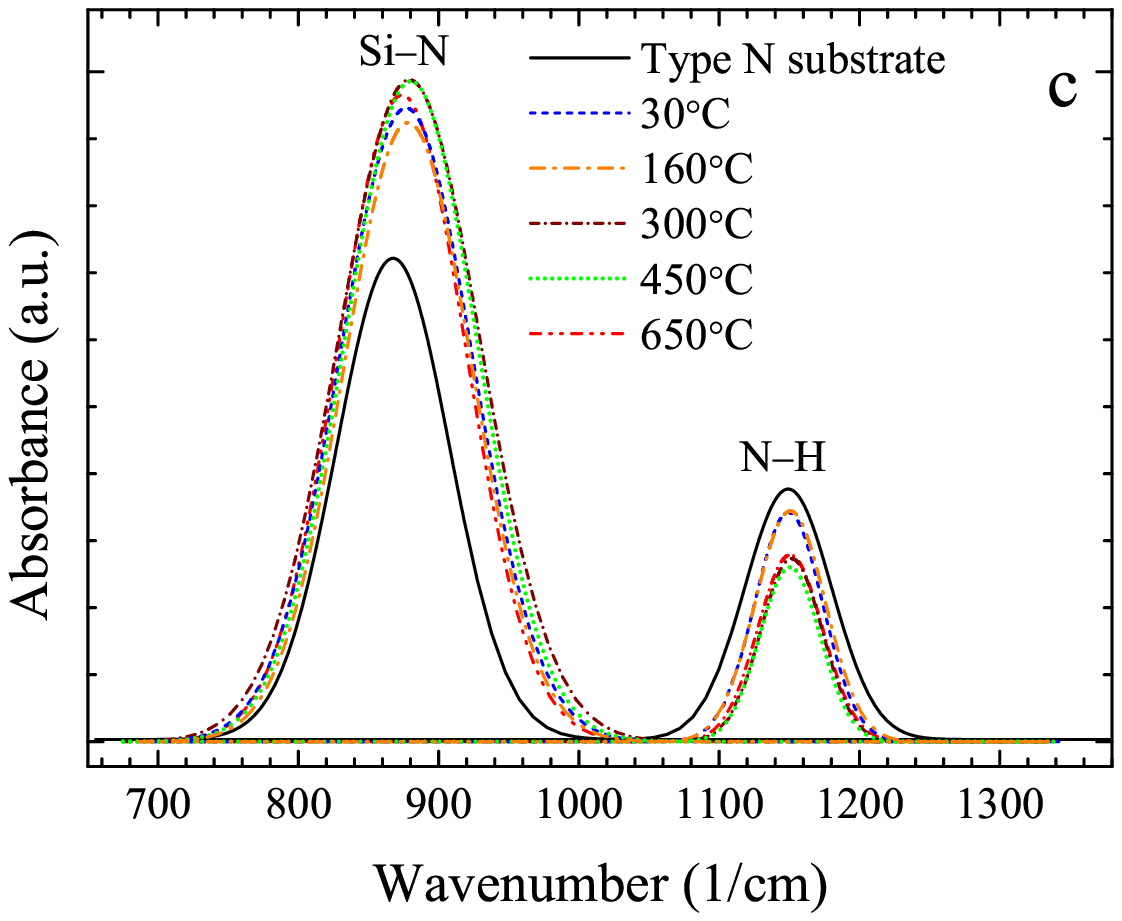}%
	\caption{FTIR absorbance spectra for the samples grown from MB at different temperatures on Type~N substrates:
		(a) spectra obtained at a Type~N substrate and at samples grown at three different temperatures indicated in the plot, the spectra are normalized to the maximum of the peak at 1090~cm$^{-1}$;  
		(b) deconvolution of the absorbance spectral band in the range from 700 to 1350 cm$^{-1}$ (the sample grown at {30\textcelsius}): the band is composed of eight Gaussian bands assigned to vibrations of the Si--O, Si--N and N--H bonds;
		(c) evolution of the bands assigned to the vibrations of the Si--N and N--H bonds in the spectra obtained at a Type~N substrate and at samples grown at three different temperatures shown in the plot.
\label{fig:MB}	 
	}
\end{figure}

Figure~\ref{fig:MB} demonstrates IR absorbance spectra in the range from 650 to 1350 cm$^{-1}$ for the samples grown from MB at different temperatures of Type~N substrates.
The spectra obtained from a Type~N substrate and from samples grown at three different temperatures are shown in Figure~\ref{fig:MB}a
(the spectra are normalized to the maximum of the highest peak at 1090~cm$^{-1}$). 
The spectra are characterized by the presence of two wide bands peaked at 825 and 1090~cm$^{-1}$, and the latter, the strongest of them assigned to the Si--O bond vibrations, exhibits a wide shoulder on the side of large wavenumbers spreading to nearly 1300~cm$^{-1}$ that is associated to the vibrations of Si--O and N--H bonds \cite{Arrangement_Si_O,FTIR_Th_&_LPCVD_SIO2,Characterization_Si_N}.
The deconvolution of these bands using Gaussian functions and analysis of vibrational frequencies presented in the panel (b) demonstrates that they are composed of eight spectral components peaked at about 735, 805, 875, 965, 1075, 1100, 1150  and 1190~cm$^{-1}$.
The peaks at 1150 and 1190~cm$^{-1}$ are assigned to the vibrations of the N--H and Si--O bonds, respectively \cite{Arrangement_Si_O,FTIR_Th_&_LPCVD_SIO2,Characterization_Si_N}.
Two strong peaks around 1090~cm$^{-1}$ are assigned to the stretching vibrations of the Si--O bond and a peak around 800~cm$^{-1}$ is attributed to the deformation vibrations of the Si--O bond \cite{Vorsina_2011,Moore_2003,Lippincott_1958}. 
Absorption peak at 965~cm$^{-1}$ likely relates to the fundamental antisymmetric stretching vibrations of the Si$_2$O$^{4-}_6$ silicate units \cite{glass_units}.
A weak absorption peak at 735~cm$^{-1}$ is also assigned to the vibrations of the Si--O bond, whereas the band around 875~cm$^{-1}$ relates to the absorption of radiation by the Si--N bond \cite{Characterization_Si_N,H_content_plasma-SiN}.

Comparison of spectral peaks plotted in Figure~\ref{fig:MB}c shows that decrease in the intensity of the absorption band relating to the vibrations of the N--H bond and simultaneous increase in the intensity of the absorption band attributed to the vibrations of the Si--N bond occur due to the silicon layer growth.
 
The peaks associated with the N--H bond are seen to virtually match for the samples produced at the temperatures of 300, 450 and 650{\textcelsius}; 
for the samples produced at the temperatures of 30 and 160{\textcelsius} these peaks also coincide but go some higher than those from the higher temperature group. 
Only the peak recorded at Type~N substrate goes separately being the most intense.
On the contrary, the peak related to the vibrations of the Si--N bond is much less intense in the spectrum obtained at Type~N substrates than those recorded at samples with deposited Si films regardless their processing temperature; and the latter are identical within the standard errors of their parameters.
This observation shows that the additional Si--N bonds begin emerging in the silicon nitride layer as soon as the Si film starts forming, even at the temperature close to the room temperature.
This process may be caused both by hydrogen diffusion from the interface of Si nitride and Si into the growing Si film and by Si accumulation at the interface.
Breaking rate of the N--H bonds initially low at low temperatures likely becomes higher at elevated temperatures.


\begin{figure}[t]
	\includegraphics*[width=0.9\linewidth]{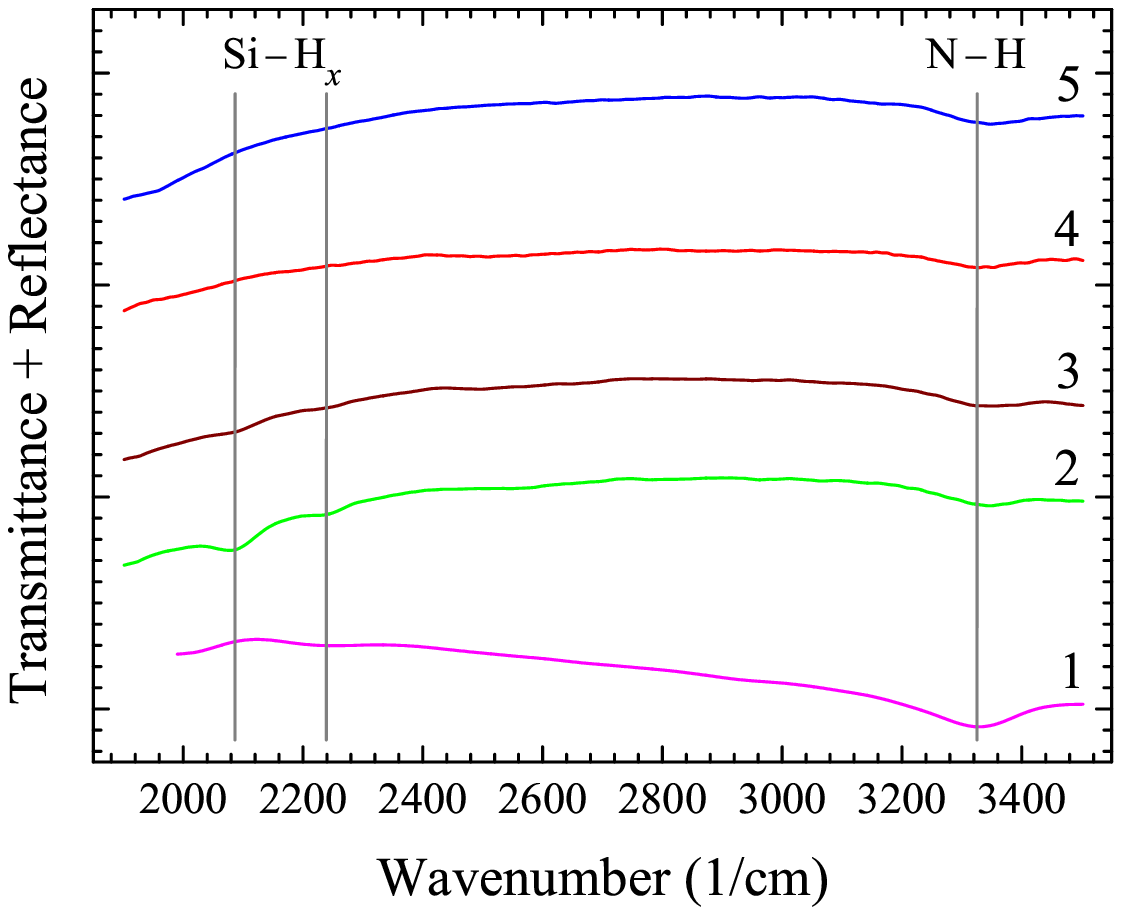}%
	\caption{Sums of FTIR transmittance and reflectance spectra for the samples grown from MB on Type~N substrates at different temperatures of silicon deposition: (1) a Type~N substrate, (2) {30\textcelsius}, (3) {160\textcelsius}, (4) {300\textcelsius}, and (5) {650\textcelsius}; the bands corresponding to the N--H and Si--H bonds vibrations are shown with vertical lines.
		\label{fig:N-H}	 
	}
\end{figure}

Figure~\ref{fig:N-H} presents the sums of IR transmittance and reflectance spectra obtained in the wavenumber range from 1900 to 3500~cm$^{-1}$ for the samples grown from MB on Type~N substrates at different temperatures of silicon deposition. 
Radiation absorption band at about 3300~cm$^{-1}$ is known to be related to the valence vibrations of the N--H bonds \cite{Characterization_Si_N,H_content_plasma-SiN}.
The intensity of this peak decreases as the silicon deposition temperature increases.
For the samples grown at low temperatures,  {30 and 160\textcelsius}, two absorption lines emerge in the wavenumber interval from 2000 to 2300~cm$^{-1}$: 
the  line peaked at about 2100~cm$^{-1}$ is connected with the Si--H bond vibration; that peaked at ca.~2250~cm$^{-1}$ presumably  may also be assigned to the Si--H bond vibration.

\begin{figure}[t]
	\includegraphics*[width=0.9\linewidth]{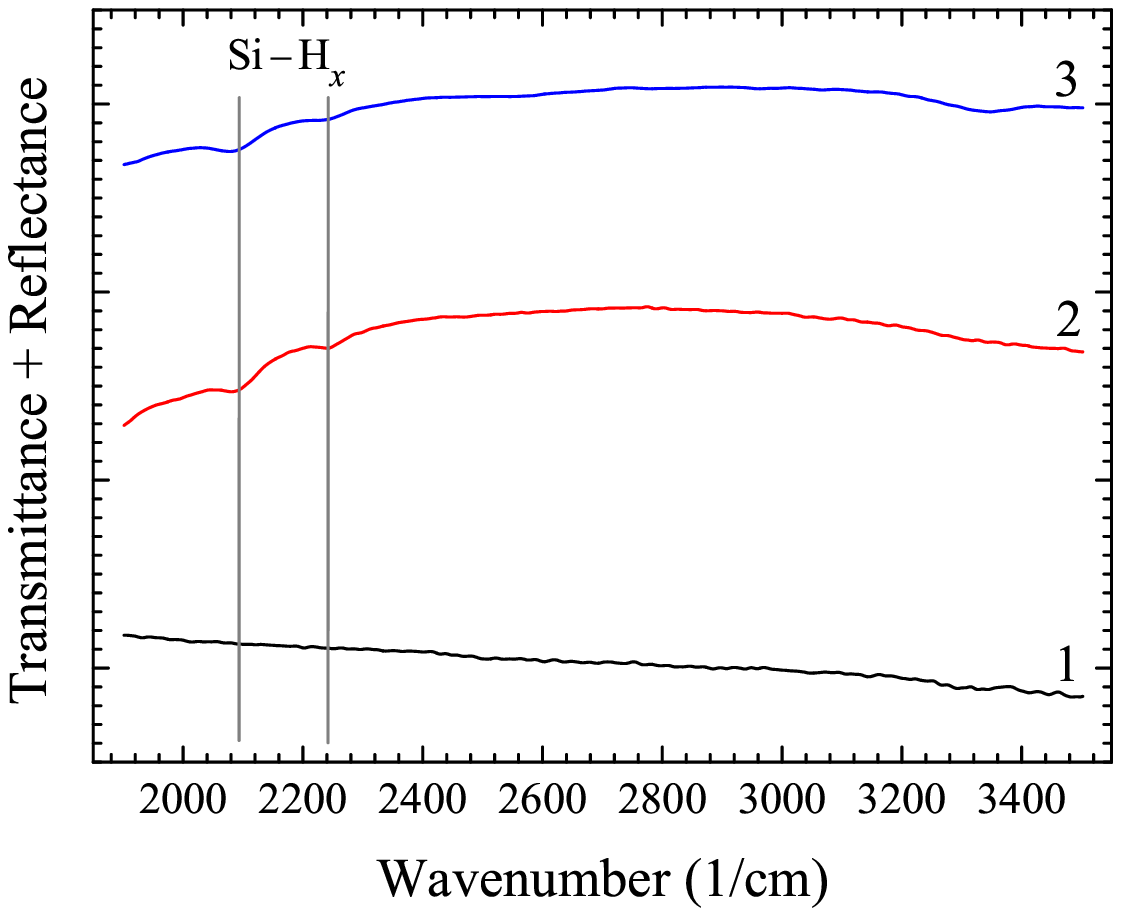}%
	\caption{Sums of FTIR transmittance and reflectance spectra for the samples grown from MB at {30\textcelsius} on (1) Type~S, (2) Type~O, and (3) Type~N substrates; the bands related to the Si--H bond vibrations are shown with vertical lines.
		\label{fig:Si-H}	 
	}
\end{figure}

Figure~\ref{fig:Si-H} depicts the sums of IR transmittance and reflectance spectra recorded in the wavenumber interval from 1900 to 3500~cm$^{-1}$ for the samples with silicon layers deposited at {30\textcelsius} from MB on different types of substrates.
It is seen that the bands peaked at $\sim2100$ and $\sim2250$~cm$^{-1}$ are present in the spectra obtained at the samples grown on Type~O substrates, but they are absent in the spectra obtained at the samples with Type~S substrates.

\begin{figure}[t]
	\includegraphics*[width=0.5\linewidth]{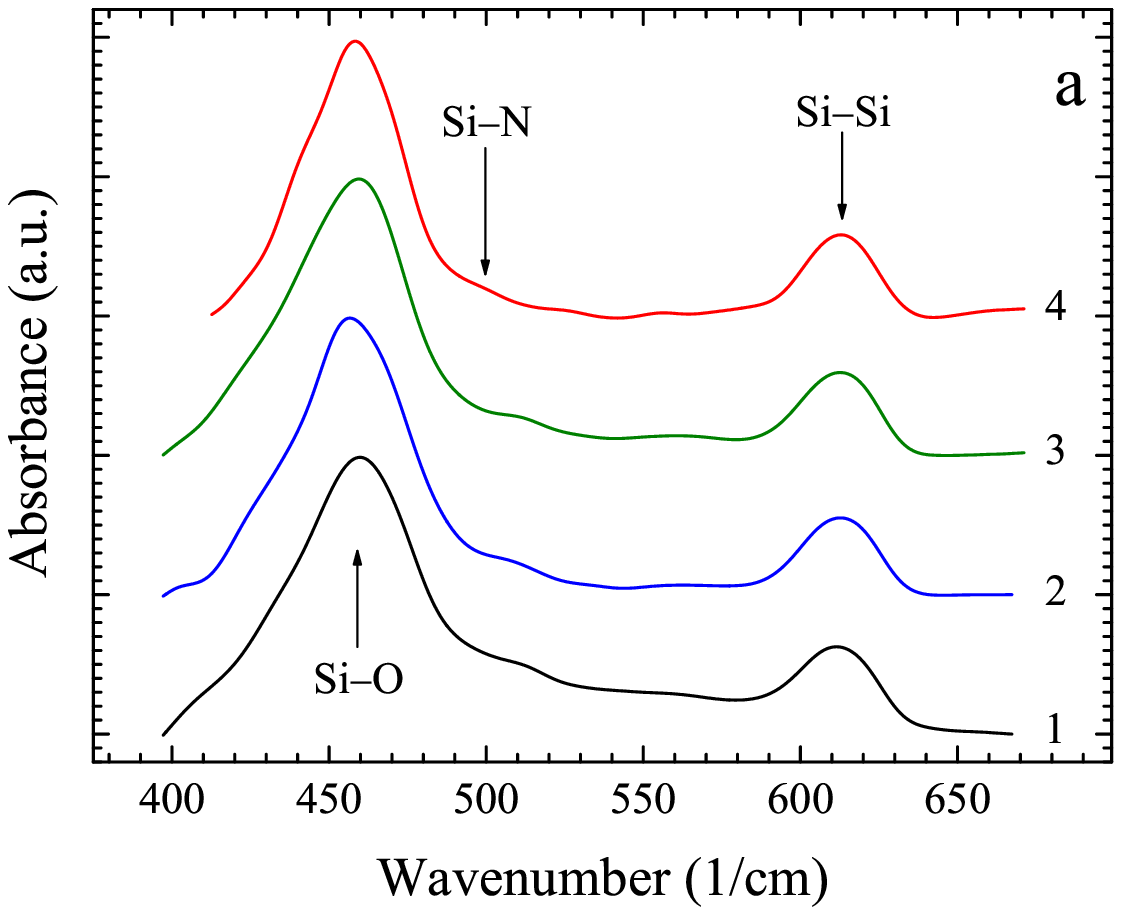}%
	\includegraphics*[width=0.5\linewidth]{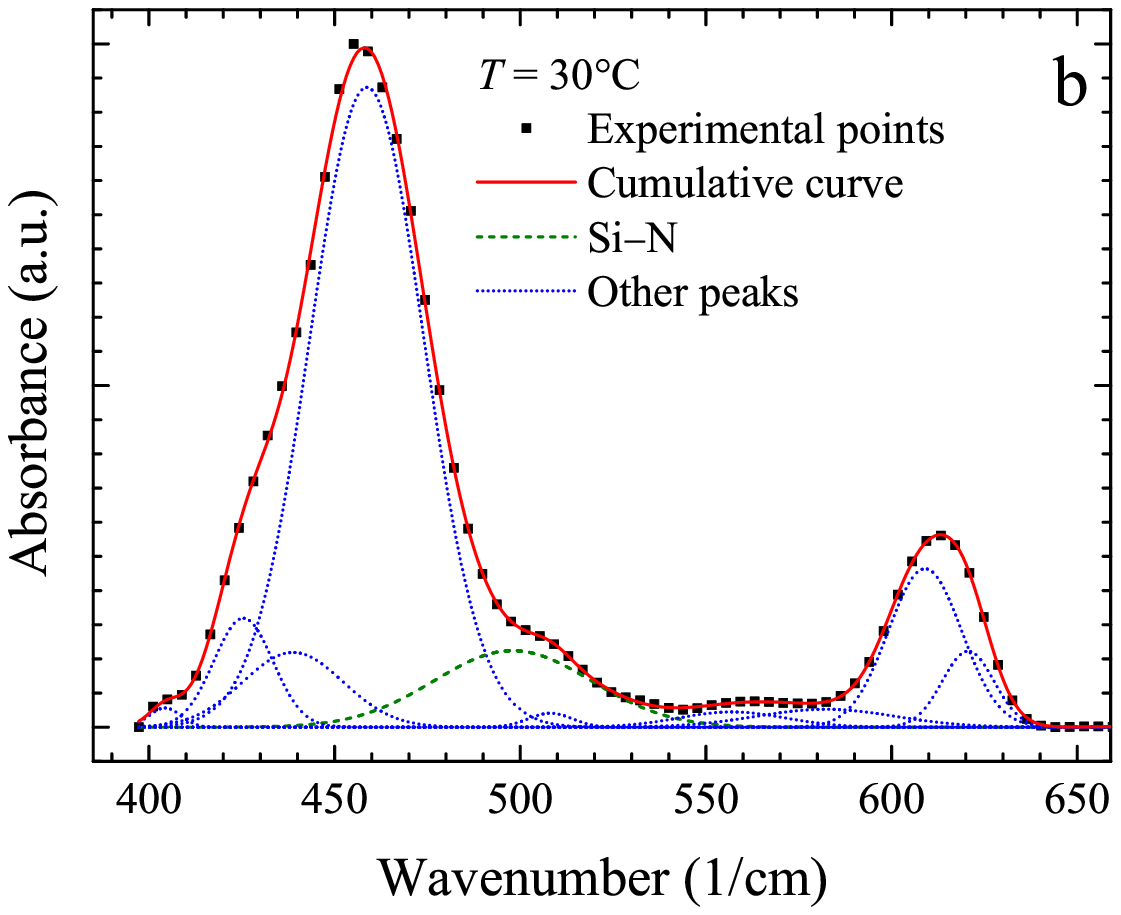}\\
	\includegraphics*[width=0.5\linewidth]{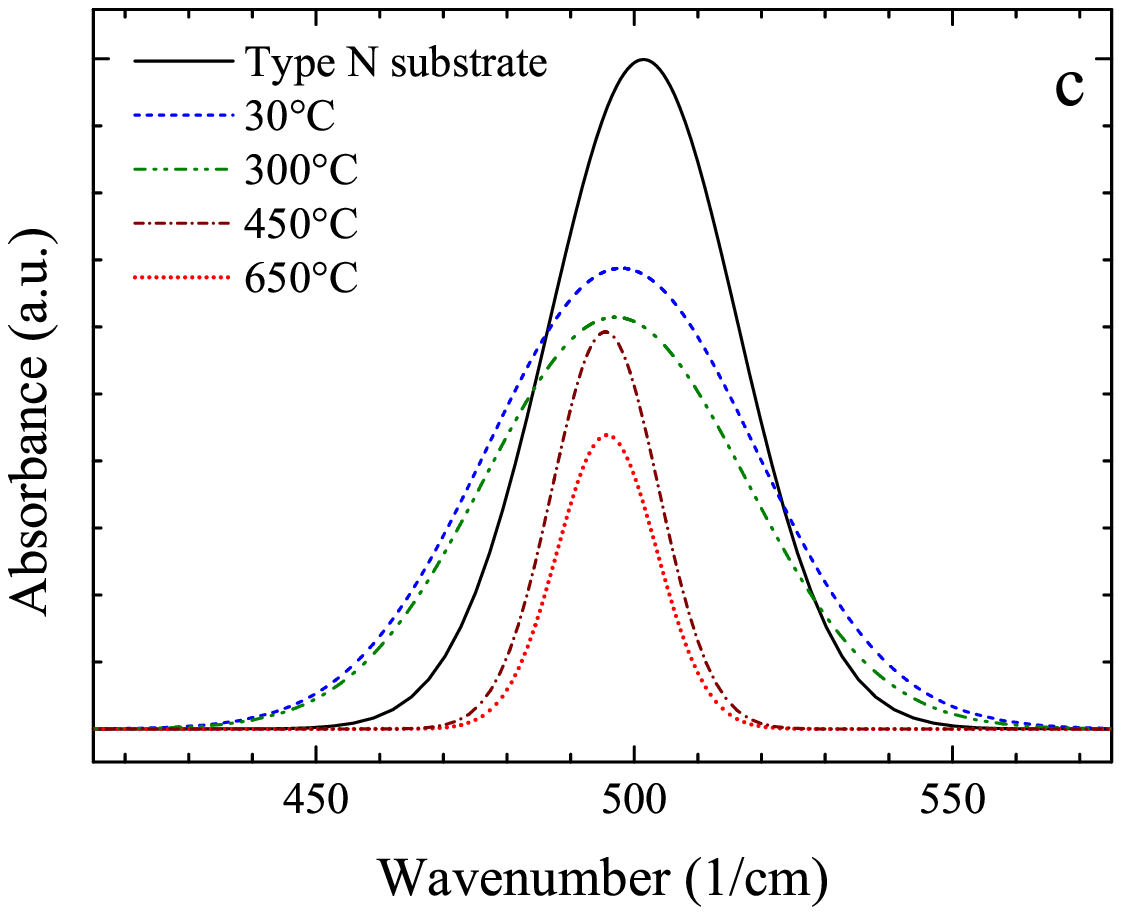}%
	\caption{%
	(a)	FTIR absorbance spectra for the samples grown from MB at different silicon deposition temperatures on Type~N substrates: (1) a Type~N substrate, (2) {30\textcelsius}, (3) {300\textcelsius}, and (4) {650\textcelsius}; the spectra are normalized to the maximum of the Si--O vibration band peaked at 458~cm$^{-1}$; the bands relating to the vibrations of the Si--O, Si--N and  Si--Si bonds are shown with vertical arrows;
	(b) deconvolution of the absorbance spectral bands in the range from 400 to 650 cm$^{-1}$ (the sample grown at {30\textcelsius}): the bands are composed of ten Gaussian peaks associated with the vibrations of the Si--O, Si--N and Si--Si bonds;
	(c) bands attributed to the vibrations of the Si--N bond in a Type~N substrate and samples grown from MB at different silicon deposition temperatures indicated in the plot.
		\label{fig:400-700}	 
	}
\end{figure}$  $

Figure~\ref{fig:400-700}a demonstrates IR absorbance spectra recorded in the range 400 to 670~cm$^{-1}$.
The absorbance curves are normalized to the maximum of the peak at $\sim 450$~cm$^{-1}$ corresponding to the vibrations of the Si--O bond.
Two characteristic bands are observed in this spectral region: a band peaked at $\sim 460$~cm$^{-1}$ and that peaked at $\sim 615$~cm$^{-1}$.
The strong wide band with a maximum at 460~cm$^{-1}$ is formed by the superposition of several absorption bands related to the vibrations of the Si--O bond with the maximums at about 400, 425, 440, 460 and 510~cm$^{-1}$ (Figure~\ref{fig:400-700}b);
the peak with the maximum around 500~cm$^{-1}$ is assigned to the vibrations of the Si--N bond. 
As the silicon deposition temperature is increased, the intensity of the absorption line at $\sim500$~cm$^{-1}$ decreases (Figure~\ref{fig:400-700}c).
Such behavior of this absorption band is associated in the literature \cite{Nitrogen-bonding} with the decrease of H atoms content in the Si$_3$N$_4$ layer and hence with the reduction of the number density of the Si--H bonds in it. This, in turn, increases an order in the  Si$_3$N$_4$ layer and decreases the absorption at $\sim500$~cm$^{-1}$ since, the more disordered silicon nitride, the higher the absorption at this line.
It should be noticed, however, that the peak at ca.~500~cm$^{-1}$ demonstrates significant narrowing after poly-Si film depositing at the temperatures higher than {450\textcelsius}, yet keeping the general trend to decrease in intensity with the growing Si deposition temperature.

The absorption band at $\sim500$~cm$^{-1}$ was not observed in the samples grown at {30\textcelsius} on Type~O and Type~S substrates.

\section{Discussion \label{discuss}}

As it is seen from the above data of our investigation of silicon layers grown on different substrates at different deposition temperatures using different deposition methods, the main changes are observed at the bands assigned to the vibrations of the N--H, Si--H and Si--N bonds.
All the Si layer deposition processes used in this work were carried out  at the temperatures lower than the deposition temperatures of the Si$_3$N$_4$ and SiO$_2$ layers.
The preliminary UHV thermal treatment at {600\textcelsius} did not change the absorption spectra of the original substrates. 
Previously, in CVD-deposited Si$_3$N$_4$ films, the reduction of the hydrogen atoms content was observed only after annealing at the temperatures that exceeded their deposition temperature \cite{Hydrogen_concentration_profiles}.
Thus, we can state that the observed changes in the IR absorption spectra are due to Si deposition.

We have found that the reduction of IR absorption at the lines related to the valence vibrations of the N--H bond at $\sim 3300$~cm$^{-1}$ and the increase in the intensity of the Si--N bond valence vibrations at $\sim 875$~cm$^{-1}$ is observed regardless of the method of depositing the Si layer on a Type~N substrate at the growth temperature in the range from 30 to {650\textcelsius}.
Additionally, the absorption band intensity of the Si--N bond vibration at $\sim 500$~cm$^{-1}$ decreases.
These changes in the absorption spectra are highly likely result from the reduction of the hydrogen content in the  Si$_3$N$_4$ layer:
hydrogen atoms diffuse into the growing amorphous or polycrystalline Si layer.
They either remain in the growing Si film or desorb from it depending on the deposition temperature.
If the latter is 30 or {160\textcelsius} that is below the hydrogen desorption temperature \cite{Hydrogen_effusion} the absorption band connected with the valence vibrations of the Si--H bond at $\sim 2100$~cm$^{-1}$ appears in the spectra.
Emergence of the band at $\sim 2250$~cm$^{-1}$ is usually connected in the literature with H--Si(O$_3$) or Si--O--Si--H structures forming in SiO$_x$:H layers \cite{Vibrational_properties_SiO_SiH,H_a-Si_bonding,Si_nitride_diimide,Hydrogen_bonding_optical_waveguides}.
In addition to the band of valence vibrations at $\sim 2250$~cm$^{-1}$, a band of deformation vibrations of those structures at $\sim 880$~cm$^{-1}$ is present in the samples studied in the cited articles.
In the current study, the absorption band assigned to the valence vibrations of the Si--N bond is registered in the wavenumber range around 875~cm$^{-1}$.
The increase in the absorption peak intensity at this spectral range does not correspond with the change in the intensity of the band at $\sim 2250$~cm$^{-1}$.
The accumulation of hydrogen atoms in the amorphous Si film growing at 30{\textcelsius} on a Type~O substrate also manifests itself in the appearance of the absorption band at $\sim 2100$~cm$^{-1}$;
the band at $\sim 2250$~cm$^{-1}$ also appears in this case, but no increase in the absorption intensity at $\sim 875$~cm$^{-1}$ is observed.
Thus, the origin of the absorption band at $\sim 2250$~cm$^{-1}$ is presumably also related to the vibrations of the Si--H bond since the conditions for its emergence are the same as for the band peaked at $\sim 2100$~cm$^{-1}$.
During Si deposition on a Type~O substrate, the SiO$_2$ layer is a source of hydrogen atoms, which are known to be present in pretty high concentration in SiO$_2$ layers obtained using thermal oxidation.
These bands were absent in the IR absorption spectra obtained at the reference sample grown at {30\textcelsius} on a Type~S substrate.

The observed decrease of intensity of the absorption band peaked at $\sim 500$~cm$^{-1}$ may also be related to the reduction of hydrogen atoms concentration in the Si$_3$N$_4$ layer. 
For instance, in Ref.\,\cite{Nitrogen-bonding}, the activity of the band at $\sim 495$~cm$^{-1}$ has been connected with the N--Si--H vibrations in a PECVD Si$_3$N$_4$ layer where hydrogen atoms form Si--H and N--H bonds and the ratio of concentrations of these bonds depends on the composition of the reacting mixture.
According to that work, reduction of the N--Si--H bonds quantity due to breaking Si--H bonds results in absorption lowering at the band at $\sim 495$~cm$^{-1}$ and in order improvement in the arrangement of atoms.
It may be supposed following that work that the observed change in the intensity of the band at $\sim 500$~cm$^{-1}$ is related to the order improvement in the Si$_3$N$_4$ layer.
In the case of CVD-deposited Si$_3$N$_4$, hydrogen atoms form N--H bonds since they have the highest binding energy.
The reduction of the concentration of hydrogen atoms and, as a consequence, of the quantity of the Si--N--H bonds leads to the enhancement of order in the arrangement of atoms and increasing of the Si--N bonds number.
As a result, the IR absorption at the band around $\sim 500$~cm$^{-1}$ decreases.

Let us consider a possible mechanism for the diffusion of hydrogen atoms from the Si$_3$N$_4$ or SiO$_2$ dielectric layer into the growing Si layer.
This process has an important feature: it occurs even at room temperature of Si deposition.
Diffusion at such a low temperature obviously is not of an activation nature.
The diffusion activation energy of hydrogen atoms, e.g., in layers of amorphous Si is about 1.5 eV \cite{Hydrogen_effusion}; 
the energy required for breaking the N--H bond is $\sim 4$~eV; the energy necessary for breaking the Si--H bond is $\sim 3$~eV \cite{Defects_hydrogen_a-SiN}.
In the discussed case, a mechanism of hydrogen atoms diffusion  may be described using a model considered in Refs.\,\cite{Hydrogen_chemical_potential,Hydrogen_migration,Silicon-Hydrogen_Bonding,Hydrogen_transport_a-Si,Diffusion_evolution_hydrogen,Hydrogen_incorporation}.
The model is based on an assumption that the diffusion of hydrogen atoms in amorphous silicon is limited by capturing of hydrogen atoms by  dangling bonds of Si atoms with formation of the Si--H bonds, which are deep traps.
Hydrogen atoms move through interstitial states that is accompanied by the rupture of the weak Si--Si bonds.
A hydrogen atom diffuses until a Si--H bond is again formed.
A concept of a hydrogen atom chemical potential ($\mu_{\mathrm H}$) is introduced in the model, the position of which is controlled by the concentration of hydrogen atoms in a layer.
For the diffusion to start, a hydrogen atom should transit from an energy level corresponding to the Si--H bond onto a so-called transport level
(to the interstitial position).
The $\mu_{\mathrm H}$ level is located between the level of deep traps and the transport level.
The greater the concentration of hydrogen atoms, the higher the $\mu_{\mathrm H}$ level.
When the layers with different levels of hydrogen atoms are combined, hydrogen atoms will diffuse from the region with a high level of $\mu_{\mathrm H}$ into the region with a low level.
In our case, the content of hydrogen atoms is higher in dielectric layers of Si$_3$N$_4$ or SiO$_2$.
The diffusion process of hydrogen atoms from the Si$_3$N$_4$ (or SiO$_2$) layer into the growing Si layer is possible until the chemical potentials of hydrogen atoms in them are equalized. 
The thickness of the Si$_3$N$_4$ layer, in which the concentration of hydrogen atoms decreases, and the depth of penetration of hydrogen atoms into the growing Si layer will expand as the deposition temperature increases.
The probability of desorption of hydrogen atoms from the growth surface of the Si layer will also increase.
As a result, at low growth temperatures, we observe the bands corresponding to the Si--H bonds in the absorption spectra, and as the temperature is raised, hydrogen atoms start to diffuse through the Si layer onto its surface and desorb \cite{Hydrogen_transport_a-Si}.
This explains the observed reduction of IR radiation absorption at the spectral bands related to the Si--H bonds at high silicon deposition temperatures. 

It should be noted also that if we take into the account a mechanical stress usually present in the studied layered structures, especially in the Si layer, we should conclude that the stress decreasing the atomic bond breaking barrier in the Si--Si bonds \cite{Atomic-Level_Fracture_Solids} will increase the hop probability of hydrogen atom during diffusion, which, as mentioned above, is accompanied with the Si--Si bond rupture. 
Thus, the weakening of the Si--Si bonds will accelerate the diffusion. 
Even at room temperature, easier Si--Si bond breaking due to local energy fluctuations will stimulate the diffusion of hydrogen atoms from a dielectric substrate into the growing silicon layer. 
The same assumption may be made about the N--H and Si--H bonds in the dielectric substrates: 
the loaded bonds should be easier to rupture and hydrogen atoms should be easier to release and hence easier to start their diffusion into the silicon film.
The internal stress, due to, e.g., difference in temperature coefficients of expansion of layers, and hence atomic bond loading is certainly extremely temperature sensitive.
The local energy fluctuations may be formally considered as a sort of  quasi-particles that take part in the diffusion process by breaking atomic bonds.
The density and the magnitude of the local energy fluctuations, as well as their movement in solid layers, also critically depend on temperature. 
So, we should consider the mechanical strain in the layers and the local energy fluctuations as additional factor influencing the process of the hydrogen atoms diffusion in silicon films deposited on dielectric substrates.

\section{Conclusion \label{concl}}

Concluding the article we emphasize its main inference:
the growth of a silicon film on a Si$_3$N$_4$ layer deposited using CVD or on a SiO$_2$ layer formed using thermal oxidation of a silicon substrate is accompanied by the diffusion of hydrogen atoms from the dielectric layer into the growing film.
In a wide temperature range, the intensity of the IR absorption bands related to the vibrations of the N--H bonds reduces whereas the intensity of the IR absorption bands connected with the Si--N bonds vibrations increases that demonstrates the escape of hydrogen atoms from a Si$_3$N$_4$ layer. 
At Si deposition temperatures lower than the temperature of hydrogen atoms desorption from the Si surface, the  absorption band assigned to the valence vibrations of the Si--H bond peaked at $\sim 2100$~cm$^{-1}$ emerges in IR spectra obtained at samples deposited both on Si$_3$N$_4$ and SiO$_2$ dielectric layers that demonstrates the accumulation of hydrogen atoms in silicon films.

The experimental results may be explained using a model of hydrogen atoms diffusion proposed in Ref.\,\cite{Hydrogen_chemical_potential}.
The migration of hydrogen atoms from the Si$_3$N$_4$ or SiO$_2$ layer into the growing silicon film is due to the difference in chemical potentials of hydrogen atoms in the dielectric layer and the silicon film.

\section*{Acknowledgments}

We thank Mr. Oleg Y. Nalivaiko of JSC Integral for the Type~N and Type~O substrates. 
We also thank the Center for Collective Use of Scientific Equipment of GPI RAS for the support of this research via presenting admittance to its equipment.



\bibliography{Literature_on_Si-Ge}

\end{document}